%% file: paper.tex
\documentclass[sigconf]{acmart}

\usepackage{booktabs} 

\usepackage{colortbl}

\usepackage{tikz}
\usetikzlibrary{shapes,arrows,positioning,decorations.markings,chains,fit,shapes.multipart}
\usepackage{import}
\usepackage{times}
\usepackage{graphicx}
\usepackage{amsfonts}
\usepackage{amsmath}
\usepackage{amssymb}
\usepackage{rotating}
\usepackage{color}
\usepackage{listings}
\usepackage{verbatim}
\usepackage{comment}
\usepackage{alltt}
\usepackage{mdwlist}
\usepackage{multirow}
\usepackage[algo2e,linesnumbered,ruled,lined]{algorithm2e}
\usepackage{url}

\makeatother

\newcommand{\tool}[1]{\textsc{#1}\xspace}

\newcommand{\cbmcv}{\tool{cbmc}}

\newcommand{\hwcbmcv}{\tool{hw-cbmc}}

\newcommand{\verifox}{\tool{CoVerIf}}

\newcommand{\klee}{\tool{klee}}

\newcommand{\Omit}[1]{}

\input{symbols}

\input{refs}

\lstdefinestyle{base}{
  language=C,
  emptylines=1,
  breaklines=true,
  basicstyle=\ttfamily\color{black},
  moredelim=**[is][\color{red}]{@}{@},
}

\title{Hardware/Software Co-verification Using Path-based Symbolic Execution} 

\author{Rajdeep Mukherjee}
\affiliation{
 \institution{Cadence Design Systems, USA}
 }
 \email{mrajdeep@cadence.com}
\author{Saurabh Joshi}
  \affiliation{
  \institution{IIT, Hyderabad}
  }
  \email{sbjoshi@iith.ac.in}
\author{John O'Leary}
  \affiliation{
  \institution{Intel Corporation, USA} 
  }
  \email{john.w.oleary@intel.com}
\author{Daniel Kroening}
  \affiliation{
  \institution{University of Oxford, UK}
  }
  \email{kroening@cs.ox.ac.uk}
\author{Tom Melham}
  \affiliation{
  \institution{University of Oxford, UK}
  }
  \email{tom.melham@cs.ox.ac.uk}
\begin{abstract}
\Omit{
Automatic verification of Hardware/Software co-designs is indispensable for dealing with 
the ever-increasing complexity of the firmware and hardware components in modern embedded systems.  
Formal methods can deliver a proof that the hardware and the firmware behave correctly, but traditional
formal methods are difficult to scale up.  
}
Conventional tools for formal hardware/software 
co-verification use bounded model checking techniques to construct a single monolithic 
propositional formula. Formulas generated in this way are extremely complex and contain a great deal of 
irrelevant logic, hence are difficult to solve even by the state-of-the-art Satisfiability (SAT) solvers.  
In a typical hardware/software co-design the firmware only exercises 
a fraction of the hardware state-space, and we can use this observation to generate simpler and more 
concise formulas. In this paper, we present a novel verification algorithm for hardware/software 
co-designs that identify partitions of the firmware and the hardware logic pertaining to the feasible 
execution paths by means of path-based symbolic simulation with custom path-pruning, property-guided 
slicing and incremental SAT solving.  We have implemented this approach in our tool \verifox.  We have 
experimentally compared \verifox with \hwcbmcv, a monolithic BMC based co-verification 
tool, and observed an average speed-up of 5$\times$ over \hwcbmcv 
for proving safety properties as well as detecting critical co-design bugs in an open-source 
Universal Asynchronous Receiver Transmitter design and a large SoC design.
\end{abstract}            

\begin{document}
\settopmatter{printacmref=false} 
\renewcommand\footnotetextcopyrightpermission[1]{} 
\pagestyle{plain} 
\setcopyright{none}

\maketitle
%
\input{introduction}
%
\input{example}
\input{methodology}
%
\input{experiment}

%
\input{background}
%
\input{conclusion}

\bibliographystyle{abbrv}
\bibliography{paper} 

\end{document}

%% file: symbols.tex
\newcommand{\safe}{\ensuremath{\mathit{Safe}}}

 \SetKwSty{fpfont}
 \SetFuncSty{fpfn}

 \SetKwFunction{Decide}{decide}
 \SetKwFunction{TransProp}{tprop}
 \SetKwFunction{UnitProp}{uprop}
 \SetKwFunction{Refines}{refines}
 \SetKwFunction{Deduce}{deduce}
 \SetKwFunction{IsUnit}{is\_unit}
 \SetKwFunction{Propagate}{propagate}
 \SetKwFunction{Invariant}{invariant}
 \SetKwFunction{Atomic}{atomic}
 \SetKwFunction{Learn}{learn}
 \SetKwFunction{Backtrack}{backtrack}
 \SetKwFunction{Clausegen}{clauseGen}
 \SetKwFunction{Invariantgen}{invariantGen}

 \SetKw{Let}{let}

\SetKwFunction{Asserting}{asserting}
\SetKwFunction{Cutheur}{cutheur}
\SetKwFunction{Generalise}{generalise}
\SetKwFunction{Analyse}{analyse}
\SetKwFunction{ClauseGeneralise}{clauseGeneralise}
\SetKwFunction{WpGeneralise}{wpGeneralise}
\SetKw{MyCase}{case}
\SetKwFunction{SetReason}{setReason}
\SetKwFunction{PickAssertingCut}{assertingCut}

\newcommand{\tuple}[1]{\ensuremath{\langle #1 \rangle}}

%% file: refs.tex
\renewcommand{\eqref}[1]{Eqn.~(\ref{Eq:#1})}

\newcommand{\algref}[1]{Alg.~\ref{Alg:#1}}
\newcommand{\algline}[1]{(Line~\ref{#1})}
\newcommand{\algtwolines}[2]{(Line~\ref{#1} and~\ref{#2})}

%% file: introduction.tex
\section{Introduction}~\label{intro}
%

In modern embedded system development, software and hardware
components are designed and implemented in parallel.
Hardware/software \emph{co-verification} is performed throughout the
design cycle to ensure that both components work
correctly together.

Before \emph{Register Transfer Level} (RTL) code exists for the
hardware components, engineers write abstract models of the proposed
hardware; such models are commonly known as {\em Transaction Level
  Model}s (TLM)~\cite{codes14}.  TLMs are typically implemented using
the SystemC TLM library~\cite{tlm} or as plain C programs.  TLMs
capture enough functionality of the hardware (HW) to enable executing
and debugging of the software (SW) or firmware (FW) before the RTL is
available~\cite{codes14, codes15}, but TLMs are always
incomplete. Co-verification of the TLM and the SW is typically
performed by testing~\cite{hvc,codes14}. We use the
  term FW and SW interchangeably in this paper.

Once RTL coding for the hardware components is complete (that is,
post-RTL), hardware/software co-verification becomes more complex.
Unlike the TLM, which only captures limited design functionality, the
RTL code describes the cycle-accurate behavior of the final HW, and
contains many extra-functional artefacts related to power, area, and
timing.  Because of the RTL's detail and complexity, the effectiveness
of testing is severely limited in post-RTL co-verification. Formal
verification is mandatory to ensure correctness.  Note that
  whenever we refer to ``hardware'' from this point onwards, we mean
  an RTL implementation and not a TLM.

The verification of SW written in C/C++ together with RTL coded in
Verilog/VHDL is very challenging.  First, there is a timing mismatch
between the synchronous clock-driven HW model and asynchronous
event-driven SW model. For example, the FW running on a processor
could be much faster or slower than the HW model it interacts
with. Second, there are no standard languages or techniques for
specifying properties of HW/SW co-designs. Third, the hardware is
highly concurrent and the software are frequently multi-threaded;
leading to a large number of event interleavings to analyze.  Finally,
there are few automated formal HW/SW co-verification tools that
support co-designs implemented in C/C++ and Verilog.  

Recently, Mukherjee et al.~\cite{DBLP:conf/dac/2017} presented a
formal HW/SW co-verification tool, called \hwcbmcv, that constructs a
combined HW-SW model through in-tandem symbolic execution of the SW
and the RTL code. SAT-based Bounded Model Checking (BMC) ~\cite{biere}
is used to prove safety properties of the combined model.  We will
refer to the combined HW-SW model as the \emph{co-verification model}.

In this paper, we build on the observation that monolithic BMC of the
co-verification model leads to propositional SAT formulas containing
much irrelevant logic. The size and complexity of the formulas pose
difficulties for SAT solvers, making the approach ineffective for
practical co-verification problems.
An effective and practical co-verification solution must reason only
about ``relevant" interactions between the SW and HW.  The notion of
relevance stems from a few sources: 1) the property or
\emph{co-specification} to be proved, 2) the behavior of the software, and
3) environmental assumptions.
First, the co-specification in a HW/SW co-design captures the design
intent that is to be verified.  A scalable HW/SW co-verification tool
need only check those parts of the co-verification model that pertain
to the given co-specification model.
Second, the SW in a typical HW/SW co-design only exercises a fragment
of the HW
state-space~\cite{polig2014micro,polig2014fpl,giefers2015accelerating}.
Formal tools may use this fact to verify only the HW functionality
exercised by the SW, ignoring or abstracting the rest.  This approach
can generate much simpler and concise formulas than those arising in
monolithic BMC -- most importantly, formulas that are more readily
solved by state-of-the-art SAT solvers.
Finally, assumptions about the environment of a HW/SW co-design may be
exploited to further constrain the verification state-space.
%

In this paper, we present a novel verification algorithm for HW/SW
co-designs that identifies partitions of the SW and the HW logic
pertaining to the feasible execution paths by means of path-based
symbolic execution with custom path-pruning, property-driven slicing,
and incremental SAT solving (see Section~\ref{ap1}).  We employ these
techniques in our tool \verifox to demonstrate that the
domain-specific optimizations in \verifox lead to more scalable
reasoning for HW/SW co-designs compared to \hwcbmcv.
%

%
\subsection{Contributions}
%
In this paper, we present a novel verification algorithm for HW/SW
co-designs called, \verifox, using path-based symbolic simulation with
custom path-pruning, property-guided slicing, and incremental SAT solving
techniques.  \verifox supports HW designs in Verilog RTL (IEEE SystemVerilog
2005 standards) and SW in ANSI-C (C89, C99 standards).  We experimentally
compare two approaches for formal HW/SW co-verification -- 1) the
\emph{monolithic} approach used in \hwcbmcv, and 2) the \emph{path-based}
approach of \verifox.  We study an open-source Universal Asynchronous
Receiver Transmitter (UART) design and a large SoC design, and we find that
\verifox is 5$\times$ faster than \hwcbmcv for proving safety properties as
well as for detecting critical co-design bugs.
%

%% file: example.tex
\section{Working Example}~\label{example}
%
Figure~\ref{firmware} gives a fragment of a SW driver for a UART design.  
The main module of the UART SW, shown on the right side of Figure~\ref{firmware}, begins 
by resetting the UART HW which is followed by a $wb\_idle()$ function (explained next).  
 The SW implements Linux style $inb()$ and $outb()$ functions which
 further invoke the $wb()$ class of functions to communicate with the
 UART HW.
The SW then configures the UART in {\em loopback} mode using $outb()$ function calls. 
The $wb()$ class of \textit{interface functions}, shown on the left side 
of Figure~\ref{firmware}, communicate with the {\em wishbone} bus interface 
to set/reset (wiggle) the UART input ports and read/write data through the 
bus interface. The calls to the top-level UART module is represented by 
$UART\_top()$ function. 
We verify that the transmitted data is the same as the 
received data in the {\em loopback} mode.  To do so, we place an assertion 
given by the $assert()$ statement (marked in red) inside the main logic of the UART SW, 
on the right-side of Figure~\ref{firmware}.  
\lstdefinestyle{base}{
  language=C,
  emptylines=1,
  breaklines=true,
  basicstyle=\ttfamily\color{black},
  moredelim=**[is][\color{red}]{@}{@},
}
\begin{figure}[t]
\scriptsize
\begin{tabular}{|l|l|}
\hline
Wishbone Interface & Main Module \\
\hline
\begin{lstlisting}[mathescape=true,language=C, style=base]
typedef unsigned char u8;
unsigned char inb 
 (unsigned long port) {
  return wb_read(port);
}
void outb (u8 value, 
 unsigned long port) {
  wb_write(port, value);
}
void wb_reset(void) {
 rst_i = 1;
 UART_top();
 rst_i = 0;
 stb_i = 0; 
 cyc_i = 0;
}
void wb_idle() {
  UART_top();
}
void wb_write(_u32 addr, 
     _u8 b) {
 adr_i = addr;
 dat_i = b;
 we_i = 1;
 cyc_i = 1;
 stb_i = 1;
 UART_top();
 we_i = 0;
 cyc_i = 0;
 stb_i = 0;
}
\end{lstlisting}
&
\begin{lstlisting}[mathescape=true,language=C, style=base]
int main() {
wb_reset();
wb_idle();
// Configure the uart
outb (0x13, UART_MC);  
outb (0x80, UART_CM3);
outb (0x00, UART_CM2);
outb (0x00, UART_CM1);
outb (0x00, UART_CR);  
outb (0x03, UART_IE); 
char tx_b[] = "Hello world";
_u8 status = 0;
char rx_b[100];
int i=0,c=0,d=0;
// data transfer in loopback
for (i=0; i<1990; i++){
if (irq_o){
 status=inb(UART_IS)&0x0c;
 if(istatus==0x0c){
 //it was a tx_empty interrupt
 outb(*(tx_b+c),UART_TR); c++;
 }
 else{ //status==0x04
 //it was an rx_data interrupt
  rx_b[d] = inb(UART_TR); d++;
 }
}else {
 // no interrupt. 
 wb_idle();
 wb_idle();
 }
}
// property
@for(i=0; i<=10; i++)@
@assert(rx_b[i] == tx_b[i]);@
}
\end{lstlisting}
\\
\hline
\end{tabular}
\caption{Software driver of UART }
\label{firmware} 
\end{figure}

%% file: methodology.tex
\section{Proposed Methodology}~\label{ap1}
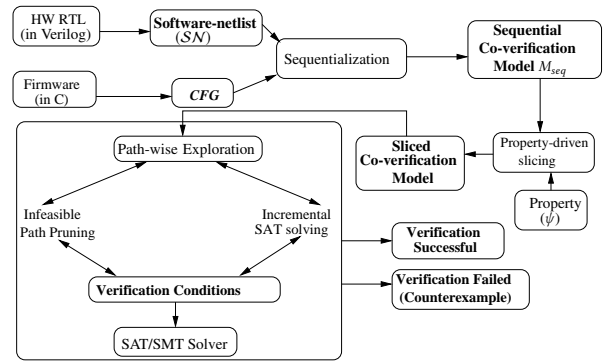
\begin{figure}[t]
\scalebox{.48}{\import{figures/}{work-flow.pspdftex}}
\caption{\small Path-based HW/SW Co-verification Flow in \verifox
\label{proposed-flow}}
\end{figure}

Figure~\ref{proposed-flow} shows our proposed verification methodology, 
as implemented in \verifox.  We now describe each step in detail.\\ 
\textbf{Step 1: Generating Software Netlist from HW}
A HW circuit, given in Verilog RTL, is automatically synthesized into a 
cycle-accurate and bit-precise \textit{software netlist}~\cite{mtk2016,mskm2016,mkm2015} 
model following synthesis semantics.  The software netlist model is  
represented as a C program~\footnote{http://www.cprover.org/hardware/v2c/} which 
retains the word-level structure as well as module hierarchy of the Verilog RTL. 
In contrast with conventional RTL synthesis into a netlist, our
software netlist exists solely to facilitate hardware/software
co-verification.
Figure~\ref{ex1} shows an example of Verilog RTL circuit 
containing both sequential and combinational elements. 
Column~2,3 in Figure~\ref{ex1} shows the formal semantics 
of the Verilog RTL and the synthesized HW respectively.  
The equivalent software netlist model is shown in column~4.

\begin{figure*}[t]
\scriptsize  
\centering
\begin{tabular}{|l|l|l|l|}
\hline
  Verilog & Formal Semantics & Synthesized Hardware & Software netlist \\
\hline
\begin{lstlisting}[mathescape=true,language=Verilog]
module top(clk, a);
input clk, a;
reg b,d,e; 
wire c,cond;
assign c = e ? 1'b0:d;
assign cond = a;
always @(posedge clk) 
 begin
  b<=a;
  if(cond && b)
   e<=b;
  else 
   e<=0;
  d<=c;
 end
endmodule
\end{lstlisting}
&
\begin{minipage}{4.2cm}
\scalebox{.5}{\import{figures/}{semantics.pspdftex}}
\end{minipage}
&
\begin{minipage}{4.0cm}
\centering
\scalebox{.5}{\import{figures/}{ckt.pspdftex}}
\end{minipage}
&
\begin{lstlisting}[mathescape=true,language=C]
struct state_elements_top {
 unsigned int b, d, e; };
struct state_elements_top  u1;
void top(_Bool clk, unsigned a) {
  _Bool c,cond;
  _Bool b_old=u1.b, d_old=u1.d;
  _Bool e_old=u1.e;
  cond = a;
  c = (u1.e)?0:u1.d;
  u1.b = a;
  if(cond && b_old)
    u1.e = b_old;
  else
    u1.e = 0;
  u1.d = c;  
}
\end{lstlisting}
\\
\hline
\end{tabular}
\caption{Circuit to Software Netlist}
\label{ex1}
\end{figure*}

\textbf{Step 2: Sequentializing Interactions of Firmware and Software Netlist}
%
%
Concurrency is a key problem for co-verification.  Testing concurrent
threads requires exploring all possible interleavings between HW and
SW threads.  The number of interleavings could potentially be
exponential.  However, we observe a specific interaction pattern,
which resembles a producer-consumer relationship.  That is, a FW
thread is mostly independent of the HW thread it interacts
with~\cite{hvc,codes14}.  Specifically, a FW thread is only
responsible for configuring the HW by writing to memory
mapped registers, or polling the interrupt status register for data
transmission, or receiving incoming data packets.
Furthermore, we observe producer-consumer interaction patterns
in many practical industrial co-designs from 
IBM~\cite{polig2014micro,polig2014fpl,giefers2015accelerating}, 
RockBox Media Player~\cite{hvc}, and others co-designs. 
In this paper, we verify co-designs that exhibit producer-consumer 
interaction behavior. 
A \textit{co-verification model}, $\mathcal{M}_{seq}$, is constructed 
through sequential composition of the FW and its interacting software netlist.

\textbf{Step 4: Property Driven Slicing of Co-verification Model}
A property-driven slicing is performed on the unified co-verification model,
$\mathcal{M}_{seq}$.  This step is purely syntactic, meaning that we perform
a backward dependency analysis starting from the property which only
preserve those program fragments that are relevant to a given property.  The
sliced program is then passed to the symbolic execution engine for
path-based exploration.  \\
%
\input{verifox}
\section{\hwcbmcv: Monolithic HW/SW Co-verification}
We briefly describe the working of the HW-SW co-verification tool, \hwcbmcv.  
In contrast to the path-based approach, in \hwcbmcv, the symbolic execution 
of HW and SW models are clearly separate and the two flows meet only at the
solver phase, where a complex monolithic formula is generated in 
Conjunctive Normal Form (CNF) from the HW and SW designs.  This complex formula 
is passed on to the solver for verification purpose.  
Furthermore, \hwcbmcv provides specific handshake primitives such
as $next\_timeframe()$ and $set\_inputs()$ to model FW-HW communications.
%
\section{Properties}\label{property}
%
%
Lack of support for property specifications in a HW/SW co-design is
one of the stumbling blocks for the application of formal techniques
in co-verification.  We express a co-design property in C language, 
which is discussed next.    

Figure~\ref{figure:component} shows an example of a temporal property for the UART HW. 
The System Verilog Assertion (SVA) is shown on the left and the equivalent property in 
C semantics is shown on the right. The temporal delay $(\#\#2)$ of the SVA on the left 
is simulated by the calls to the top-level UART module in the software netlist, which 
is represented by $UART\_top()$ function in the right column.
\begin{figure}[htbp]
\scriptsize
\begin{tabular}{l|l}
\hline
System Verilog Assertions  & Assertion (in C)
\\
\hline
\begin{lstlisting}[mathescape=true,language=Verilog]
P1: assert property 
 (@(posedge clk) 
 ack == 1|-> 
 (valid == 1 && 
 ##2 empty == 0)); 
\end{lstlisting}
&
\begin{lstlisting}[mathescape=true,language=C]
bool Property_P1() {
 assert(!ack || valid);
 UART_top();
 UART_top();
 assert(empty==0); }
\end{lstlisting} \\
\hline
\end{tabular}
\caption{Sample property of UART HW}
\label{figure:component}
\end{figure}
Figure~\ref{figure:transaction} shows few examples of co-design properties that specify 
the interaction between the FW and HW components of the UART design.     
In $Property\_P2()$, the SW event $outb(UART\_TR,0x0c)$ triggers the HW event 
$ack\_o$ (marked in bold) after one clock cycle. 
Whereas in $Property\_P3()$, the antecedent $tx\_empty()$ is a HW event and 
$send\_data$ is a SW event (marked in bold).  
\begin{figure}[htbp]
\scriptsize
\begin{tabular}{l|l}
\hline
\multicolumn{2}{c}{HW/SW Co-specification}
\\
\hline
\begin{lstlisting}[mathescape=true,language=C,moredelim={[is][keywordstyle]{@@}{@@}}]
bool Property_P2() { 
 if(outb(UART_TR,0x0c)){
  UART_top();
  assert(@@ ack_o==1 @@);}}
\end{lstlisting} 
&
\begin{lstlisting}[mathescape=true,language=C,moredelim={[is][keywordstyle]{@@}{@@}}]
bool Property_P3() { 
 assert(!tx_empty() || 
 (@@(send_data&0x1)==1)@@);
}
\end{lstlisting} \\ 
\hline
\end{tabular}
\caption{Properties capturing FW/HW interactions}
\label{figure:transaction}
\end{figure}
%

%% file: figures/work-flow.pspdftex
\begin{picture}(0,0)%
\includegraphics{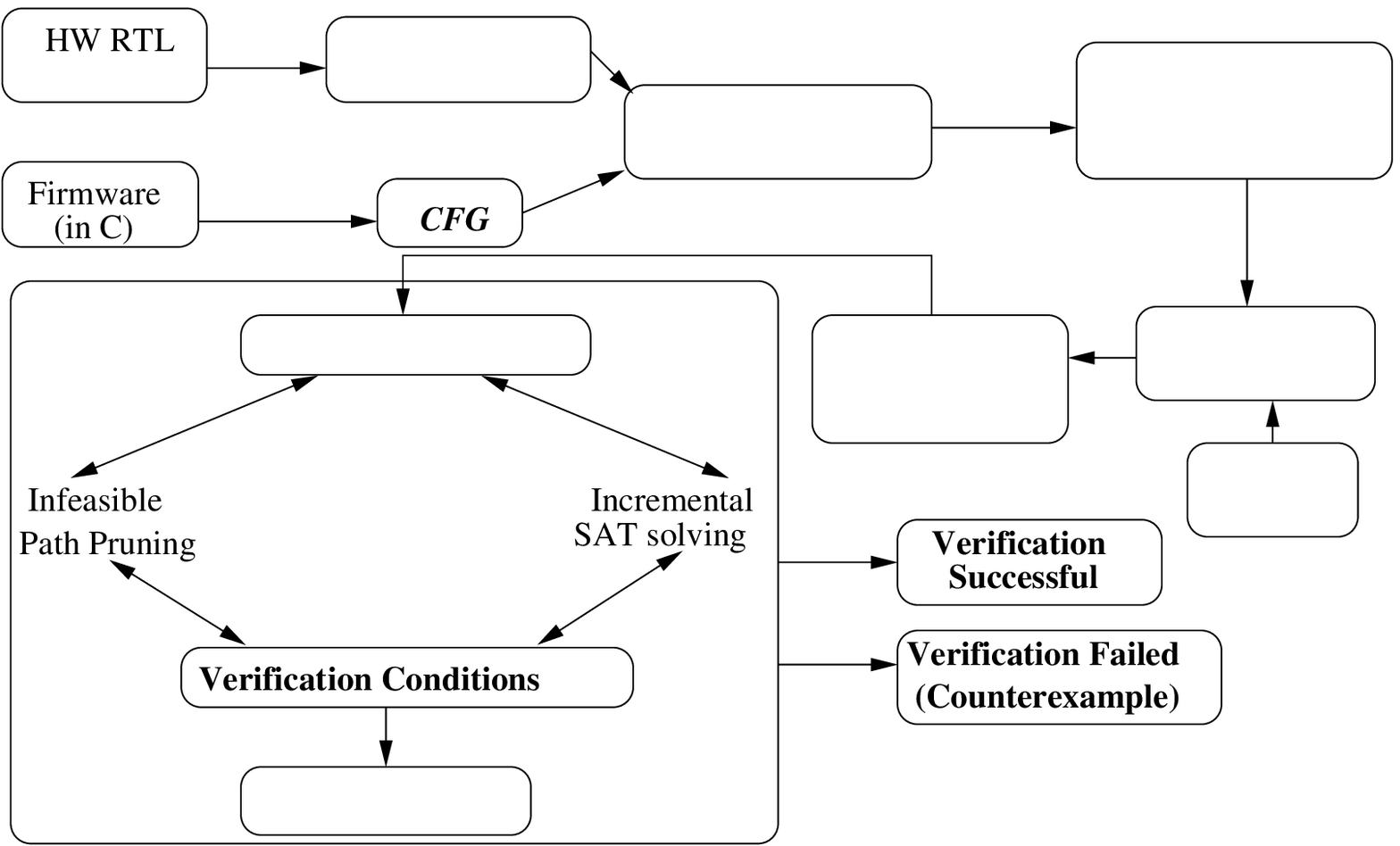}%
\end{picture}%
\setlength{\unitlength}{4144sp}%
\begingroup\makeatletter\ifx\SetFigFont\undefined%
\gdef\SetFigFont#1#2#3#4#5{%
  \reset@font\fontsize{#1}{#2pt}%
  \fontfamily{#3}\fontseries{#4}\fontshape{#5}%
  \selectfont}%
\fi\endgroup%
\begin{picture}(7359,4434)(-866,-5293)
\put(5581,-3391){\makebox(0,0)[lb]{\smash{{\SetFigFont{12}{14.4}{\rmdefault}{\mddefault}{\updefault}{\color[rgb]{0,0,0}Property}%
}}}}
\put(5716,-3571){\makebox(0,0)[lb]{\smash{{\SetFigFont{12}{14.4}{\rmdefault}{\mddefault}{\updefault}{\color[rgb]{0,0,0}($\psi$)}%
}}}}
\put(3736,-2626){\makebox(0,0)[lb]{\smash{{\SetFigFont{12}{14.4}{\rmdefault}{\bfdefault}{\updefault}{\color[rgb]{0,0,0} }%
}}}}
\put(901,-1141){\makebox(0,0)[lb]{\smash{{\SetFigFont{12}{14.4}{\rmdefault}{\bfdefault}{\updefault}{\color[rgb]{0,0,0}Software-netlist}%
}}}}
\put(1171,-1321){\makebox(0,0)[lb]{\smash{{\SetFigFont{12}{14.4}{\rmdefault}{\bfdefault}{\updefault}{\color[rgb]{0,0,0}$(\mathcal{SN})$}%
}}}}
\put(3871,-2671){\makebox(0,0)[lb]{\smash{{\SetFigFont{12}{14.4}{\rmdefault}{\bfdefault}{\updefault}{\color[rgb]{0,0,0}       Sliced }%
}}}}
\put(-764,-1276){\makebox(0,0)[lb]{\smash{{\SetFigFont{12}{14.4}{\rmdefault}{\mddefault}{\updefault}{\color[rgb]{0,0,0}(in Verilog)}%
}}}}
\put(451,-2716){\makebox(0,0)[lb]{\smash{{\SetFigFont{12}{14.4}{\rmdefault}{\mddefault}{\updefault}{\color[rgb]{0,0,0}Path-wise}%
}}}}
\put(1261,-2716){\makebox(0,0)[lb]{\smash{{\SetFigFont{12}{14.4}{\rmdefault}{\mddefault}{\updefault}{\color[rgb]{0,0,0}Exploration}%
}}}}
\put(5221,-2626){\makebox(0,0)[lb]{\smash{{\SetFigFont{11}{13.2}{\rmdefault}{\mddefault}{\updefault}{\color[rgb]{0,0,0}Property-driven }%
}}}}
\put(3871,-3076){\makebox(0,0)[lb]{\smash{{\SetFigFont{12}{14.4}{\rmdefault}{\bfdefault}{\updefault}{\color[rgb]{0,0,0}Model}%
}}}}
\put(3466,-2851){\makebox(0,0)[lb]{\smash{{\SetFigFont{12}{14.4}{\rmdefault}{\bfdefault}{\updefault}{\color[rgb]{0,0,0}Co-verification}%
}}}}
\put(496,-5146){\makebox(0,0)[lb]{\smash{{\SetFigFont{12}{14.4}{\rmdefault}{\mddefault}{\updefault}{\color[rgb]{0,0,0}SAT/SMT Solver}%
}}}}
\put(2521,-1591){\makebox(0,0)[lb]{\smash{{\SetFigFont{12}{14.4}{\rmdefault}{\mddefault}{\updefault}{\color[rgb]{0,0,0} Sequentialization}%
}}}}
\put(5446,-2851){\makebox(0,0)[lb]{\smash{{\SetFigFont{11}{13.2}{\rmdefault}{\mddefault}{\updefault}{\color[rgb]{0,0,0}     slicing}%
}}}}
\put(4951,-1456){\makebox(0,0)[lb]{\smash{{\SetFigFont{12}{14.4}{\rmdefault}{\bfdefault}{\updefault}{\color[rgb]{0,0,0}Co-verification}%
}}}}
\put(5131,-1231){\makebox(0,0)[lb]{\smash{{\SetFigFont{12}{14.4}{\rmdefault}{\bfdefault}{\updefault}{\color[rgb]{0,0,0}Sequential}%
}}}}
\put(5131,-1681){\makebox(0,0)[lb]{\smash{{\SetFigFont{12}{14.4}{\rmdefault}{\bfdefault}{\updefault}{\color[rgb]{0,0,0}Model $M_{seq}$}%
}}}}
\end{picture}%

%% file: figures/semantics.pspdftex
\begin{picture}(0,0)%
\includegraphics{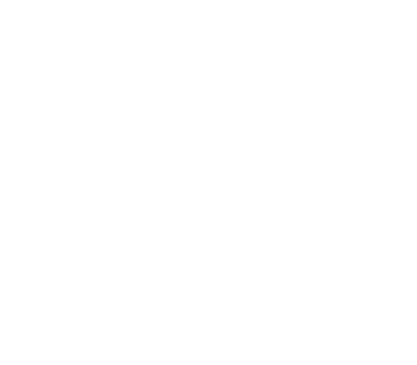}%
\end{picture}%
\setlength{\unitlength}{4144sp}%
\begingroup\makeatletter\ifx\SetFigFont\undefined%
\gdef\SetFigFont#1#2#3#4#5{%
  \reset@font\fontsize{#1}{#2pt}%
  \fontfamily{#3}\fontseries{#4}\fontshape{#5}%
  \selectfont}%
\fi\endgroup%
\begin{picture}(1891,1656)(1426,-4495)
\put(1441,-2986){\makebox(0,0)[lb]{\smash{{\SetFigFont{12}{14.4}{\rmdefault}{\bfdefault}{\updefault}{\color[rgb]{0,0,0}Combinational Logic}%
}}}}
\put(1441,-3211){\makebox(0,0)[lb]{\smash{{\SetFigFont{12}{14.4}{\rmdefault}{\mddefault}{\updefault}{\color[rgb]{0,0,0}$\forall t,\; c(t)= if\; e(t)\; then\; 0\; else\; d(t)$}%
}}}}
\put(1441,-3436){\makebox(0,0)[lb]{\smash{{\SetFigFont{12}{14.4}{\rmdefault}{\mddefault}{\updefault}{\color[rgb]{0,0,0}$\forall t,\; cond(t) = a(t)$}%
}}}}
\put(1441,-3751){\makebox(0,0)[lb]{\smash{{\SetFigFont{12}{14.4}{\rmdefault}{\bfdefault}{\updefault}{\color[rgb]{0,0,0}Sequential Logic}%
}}}}
\put(1441,-3976){\makebox(0,0)[lb]{\smash{{\SetFigFont{12}{14.4}{\rmdefault}{\mddefault}{\updefault}{\color[rgb]{0,0,0}$\forall t,\; b(t+1)=a(t)$}%
}}}}
\put(1441,-4201){\makebox(0,0)[lb]{\smash{{\SetFigFont{12}{14.4}{\rmdefault}{\mddefault}{\updefault}{\color[rgb]{0,0,0}$\forall t,\; e(t+1)=if(cond(t)\; \wedge\; b(t))\; then\; b(t)\; else\; 0$}%
}}}}
\put(1441,-4426){\makebox(0,0)[lb]{\smash{{\SetFigFont{12}{14.4}{\rmdefault}{\mddefault}{\updefault}{\color[rgb]{0,0,0}$\forall t,\; d(t+1)=c(t)$}%
}}}}
\end{picture}%

%% file: figures/ckt.pspdftex
\begin{picture}(0,0)%
\includegraphics{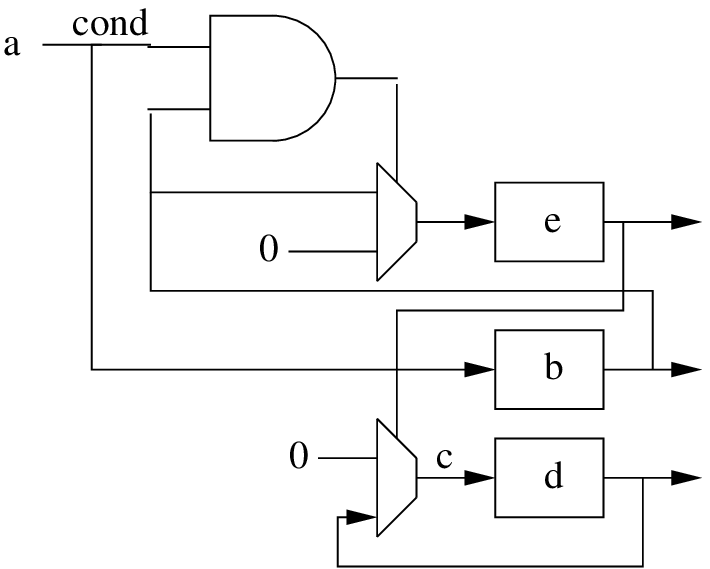}%
\end{picture}%
\setlength{\unitlength}{4144sp}%
\begingroup\makeatletter\ifx\SetFigFont\undefined%
\gdef\SetFigFont#1#2#3#4#5{%
  \reset@font\fontsize{#1}{#2pt}%
  \fontfamily{#3}\fontseries{#4}\fontshape{#5}%
  \selectfont}%
\fi\endgroup%
\begin{picture}(3222,2592)(2731,-5113)
\end{picture}%

%% file: verifox.tex
\textbf{Step 5: Co-verification Using Path-based Symbolic Execution}
Given a co-verification model, $\mathcal{M}_{seq}$, a scenario, 
$\mathcal{S}$ typically represented by assume properties, 
and a co-design property expressed as $\mathit{assert}(c)$ 
(where $c$ is a condition stated in terms of variables in $\mathcal{M}_{seq}$)
as input, \verifox performs path-based exploration of $\mathcal{M}_{seq}$ 
to automatically check its validity using backend solvers. If the condition 
$c$ does not hold, then $\mathcal{M}_{seq}$ is said to have violated the property.

A typical path-based symbolic execution engine might explore 
a path until it come to an $\mathit{assert}(c)$ statement.
This whole path can then be posed as a query to a SAT solver to see if the 
assertion is violated at that point. If the path is infeasible, the assertion
holds trivially. If a large number of paths are infeasible, symbolic 
execution may waste time exploring them. \verifox employs 
an eager infeasibility check to prune infeasible paths, 
as well as incremental encoding that makes
it easier for the underlying SAT solver. 
\begin{algorithm2e}[t]
\DontPrintSemicolon
\SetKwFunction{symex}{Symex}
\SetKwFunction{put}{put}
\SetKwFunction{get}{get}
\SetKwFunction{print}{print}
\SetKwFunction{return}{return}
\SetKwData{safe}{Safe}
\SetKwData{unsafe}{Unsafe}
\SetKwInOut{Input}{input}
\SetKwInOut{Output}{output}
\SetKw{KwNot}{not}
\begin{small}
  \Input{Co-verification Model $\mathcal{M}_{seq}$ with properties specified with $assert(c)$, scenario specified with $assume(c)$}
\Output{The status (\safe or \unsafe) and a counterexample if \unsafe}
\tcc{The initial state}
$S_0 \leftarrow I(x)$ \nllabel{algline:init}\;
$stmt \leftarrow $ first statement \;
$worklist.put(\tuple{S_0,stmt})$ \;
\While{\KwNot $worklist.empty()$}
{
	$\tuple{S,stmt} \leftarrow worklist.get()$ \;
	\uIf{$stmt$ is an $assume(c)$ \nllabel{algline:assume}}
           {
			$stmt \leftarrow $ statement after $assume(c)$ \;
			\lIf{$isFeasible(S\wedge c)$}{$worklist.put(\tuple{S\wedge c,stmt})$}
              
       		}
	\uElseIf{$stmt$ is a branch with condition $c$ \nllabel{algline:branch}}
	{
		$stmt_f \leftarrow$ first statement after $stmt$ if branch is not taken\;
		$stmt_t \leftarrow$ first statement after $stmt$ if branch is taken \;
		\lIf{$isFeasible(S\wedge c)$}{$worklist.put(\tuple{S\wedge c,stmt_t})$}
		\lIf{$isFeasible(S\wedge \neg c)$}{$worklist.put(\tuple{S\wedge \neg c,stmt_f})$}
		
	}
	\uElseIf{$stmt$ is an $assert(c)$ \nllabel{algline:assert}}
	{
		$stmt \leftarrow $ statement after $assert(c)$ \;
		\If{$isFeasible(S \wedge \neg c)$}{
		\print \unsafe \;
		\return Counterexample}
		\lElse{ $worklist.put(\tuple{S \wedge c,stmt})$}
	}
	\Else
	{
		$S \leftarrow symex(S,stmt)$\; \nllabel{algline:symex}
		$stmt \leftarrow $ the next statement in control flow after $stmt$ \;
		\lIf{$stmt \neq \bot$ \nllabel{algline:nullstmt}} {$worklist.put(\tuple{S,stmt})$} \;
		
	}
	\return \safe \;
}
\end{small}
\caption{Co-verification Using Path-based Symbolic Execution\label{Alg:verifox}}
\end{algorithm2e}

\Omit{
, in which every time a branch or an assumption
is encountered, a feasibility check is made to prune away infeasible 
paths as early as possible. }
\algref{verifox} shows the overall algorithm of \verifox. States mentioned
in the algorithm are all symbolic states, which are quantifier-free predicates characterizing
a set of program states. 
Symbolic execution starts with an initial symbolic state $I(x)$, is a quantifier-free predicate
over program variables $x$, and the first statement $\mathit{stmt}$ to be executed.  Note that we assume all program variables
have finite bit-width and thus can be represented as bit-vectors.
Every statement acts as a state transformer during the symbolic execution. $\mathit{worklist}$ 
maintains the set of symbolic states, along with the corresponding
$\mathit{stmt}$ that should execute next. Assumptions can be specified in the program 
using $\mathit{assume}(c)$ statements, where $c$ is the condition expressed in terms 
of program variables. Assumptions restrict the search to only those
states for which the condition $c$ holds at the program point where $\mathit{assume}(c)$ is encountered.
For example, suppose $(x!=0)$ characterizes the set of states that has been discovered to be
reachable so far by a verification tool. Here, $x$ is a program variable.
Upon encountering $\mathit{assume}(x>0)$, the 
set of states reachable at the point of assumption is shrunk to only those that
satisfy $(x>0)$. A user can specify assumptions to restrict
the verification to only certain regions of the program's state space.

\verifox performs a feasibility check at an $\mathit{assume}$ statement 
or a branch \algtwolines{algline:assume}{algline:branch} to ensure that
only feasible symbolic states are kept in the $\mathit{worklist}$. This ensures that the
infeasibility is detected as early as possible. 
If an assertion is violated, then a counterexample is detected and \algref{verifox}
terminates \algline{algline:assert}. In all other cases, $\mathit{symex}(S,\mathit{stmt})$ performs
one step of symbolic execution by executing $\mathit{stmt}$ from the symbolic state $S$ \algline{algline:symex}.
If no further statement remains to be executed along the path that is being explored,
then $\mathit{stmt}$ is assigned the value $\bot$. The symbolic state is put in the worklist only 
if there are further statements remaining \algline{algline:nullstmt}.

The feasibility checks shown as $\mathit{isFeasible}$ pose a query to the underlying SAT solver.
Note that \algref{verifox} does not refer to how the methods $\mathit{worklist.put}$ and $\mathit{worklist.get}$
work. In principle, one can use any search heuristic to select which symbolic state to explore 
further from the $\mathit{worklist}$. In the current version \verifox employs a depth first strategy of exploration.

Apart from the eager infeasibility check, another crucial optimization is the
use of incremental SAT solving. 
During the symbolic execution, only one solver instance is maintained 
while going down
a single path. Thus, when making a feasibility
check from one branch $b_1$ to another branch $b_2$ along a single path, only the program
segment from $b_1$ to $b_2$ is encoded as a constraint and added to the existing solver
instance. 
\Omit{
Internal solver states and the information that it gathers during the search
remains valid as long as all the queries that are posed to the solver in succession are
monotonically stronger. If the solver solves a formula $\phi$, posing $\phi \wedge \psi$ as 
a query to the same solver instance allows one to reuse the solver knowledge that
it has acquired in the previous query, because any assignment that falsifies $\phi$ also
falsifies $\phi \wedge \psi$. Thus the solver need not revisit the assignments that it has already
ruled out.}
This results in speeding up the process of feasibility check of the symbolic state at $b_2$ as
the feasibility check at $b_1$ was $\mathit{true}$. A new
solver instance is used to explore a different path, after the current path is detected as infeasible.

\Omit{
In the full incremental mode, only one solver instance is maintained throughout the whole
 symbolic execution. Let $\phi_{b_1b_2}$ denote the encoding of the path fragment
from $b_1$ to $b_2$. It is added in the solver as $(B_{b_1b_2} \implies \phi_{b_1b_2})$.
Then, $B_{b_1b_2}$ is added as a \textit{blocking variable}, which is also usually known as {\em solver assumption}.\footnote{The SAT community uses the term \textit{assumption variables} or \textit{assumptions}, but we will use the term blocking variable to avoid ambiguity with assumptions in the program.} to enforce constraints specified by $\phi_{b_1b_2}$. Blocking variables are treated specially inside the solvers: unlike regular variables or clauses, the blocking can be removed in subsequent queries without invalidating the solver instance.
When one wants to back-track the symbolic execution, the blocking $B_{b_1b_2}$ is removed 
and a unit clause $\neg B_{b_1b_2}$ is added to the solver, thus effectively removing $\phi_{b_1b_2}$.
}

\Omit{
In principle, the optimization involving incremental path encoding in Algorithm~\ref{Alg:verifox} 
works for any SAT/SMT solver.  However, we did not integrate SMT solvers with \verifox because the 
performance of incremental SMT solvers such as Z3 is very slow compared 
to the incremental SAT solvers due to the bit-manipulating nature of the HW/SW co-design.  
}

The eager infeasibility check restricts the search to explore
only those SW/HW interactions which are feasible under a
given scenario.
In our experiments, we find this optimization has a large effect on runtimes.
Though \verifox poses many queries to the SAT solver, each query is relatively simple due to
two reasons: the resultant formula encodes only a single path, 
and exploration along a path only needs to encode and solve for the path segment (along with the existing
constraints) from the last point of query.

%% file: experiment.tex
\section{Experimental Results}~\label{result}
%
We report experimental results for SW-HW co-verification of a UART  and a
SoC design.  All our experiments were performed on an Intel Xeon 3.0\,GHz
machine with 48\,GB RAM.  All times reported are in seconds.  The timeout 
for all our experiments was set to 2~hours.  The performance of bit-level 
and word-level flow in \hwcbmcv are similar. So, we only report bit-level 
results for \hwcbmcv.  MiniSAT-2.2.0~\cite{DBLP:conf/sat/EenB05} was used 
as underlying SAT solver with \hwcbmcv and \verifox.  The focus of our 
experiments is to compare the performance of \verifox against \hwcbmcv for 
verification of an UART design and a SoC design.  
We distribute our tool \verifox, along with the HW/SW co-design benchmarks 
here~\footnote{https://drive.google.com/open?id=1Y2HkfIWwf6YkJgl24OXrA-dG2rX-92bm}. 
\\ 

\textit{\textbf{Comparison with Other HW/SW Co-verification Tools:}} 
Despite extensive use of model checking and other formal methods in 
SW verification or HW verification domain, building automated 
HW/SW co-verification tools using formal methods has received 
little attention in the past.  Other than \hwcbmcv~\cite{DBLP:conf/dac/2017}, 
we are not aware of any other automated formal co-verification tool in the literature 
that can readily accept co-designs written in C/C++ and Verilog RTL.  Hence, we 
only compare our results with \hwcbmcv in this paper.  
%
\subsection{HW/SW co-verification of UART}

\textbf{\emph{About UART:}} A UART  core is used for asynchronous
transmission and reception of data which provides serial communication
capabilities with a modem or other external devices.  The UART  is
compliant with industry standards for UART and interfaces with the wishbone
bus.  The design statistics for UART is shown in Table~\ref{table:stats}. \\
\begin{table}
\begin{center}
{
\small
\begin{tabular}{|c|c|c|c|c|c|c|}
\hline
  Circuit & Verilog & Latches(L)/ & Input & Output & GATE & Firmware \\
  & LOC & FF & Ports & Ports & Count & (LOC)\\ 
\cline{1-7}
\multicolumn{7}{|c|}{Universal Asynchronous Receiver Transmitter} \\ \hline 
  UART & 1200 & 356 & 12 & 9 & 413 & 528 \\ \hline
\multicolumn{7}{|c|}{System-On-a-Chip} \\ \hline 
  SoC & 3567 & 840 & 14 & 11 & 945 & 734 \\ \hline
\end{tabular}
}
\end{center}
\caption{Design statistics for UART and SoC Design}
\label{table:stats}
\end{table}  
The UART  core is configured in 
3 different operating modes, namely--{\em transmission without interrupt
enabled} (Scenario A), {\em transmission with interrupt enabled} (Scenario
B) which transmit non-deterministic data through the serial output while 
the receiver module is inactive. In Scenario C, the UART is configured in 
{\em loopback mode with interrupt enabled} in which both the transmitter 
and the receiver are active. The data-width varies in each mode, ranging from 8~bits to 64~bits.  
\\
\textbf{\emph{Discussion of Result:}} 
Table~\ref{table:safe} reports the run times for bounded safety proofs of 
co-design properties in UART  core.  Column~1 in Table~\ref{table:safe} 
gives the name of the scenario, Column~2 gives the maximum loop unroll 
bound of the firmware, column~3-7 present the runtime using \hwcbmcv, total/feasible 
path counts, 
\verifox, respectively.  
Table~\ref{table:safe} shows that \verifox dominates \hwcbmcv 
in all scenarios (marked in bold). \verifox is on average 
$8\times$ faster than \hwcbmcv, both for proving safety as well as 
detecting bugs.  Thus, \verifox outperforms \hwcbmcv in all scenarios. \\ 
We verified a total of~39 properties of the UART design. 
Table~\ref{table:safe} reports some of the
representative properties.  In Scenario~A and scenario~B, we verified whether
the transmitted data (32-bit or 64-bit) is available through the serial
output port after a pre-determined number of clock cycles.
In both of the configurations, \verifox is able to prune the receiver logic since the
SW only exercises the transmitter module by appropriately configuring
the memory-mapped registers.  In Scenario~C, we verified whether the transmitted
data matches the data received when the UART is configured in loopback
mode. We found several bugs in the open source UART obtained 
from~\url{http://www.opencores.org}. The bottom part of 
Table~\ref{table:safe} reports the runtimes for detecting 
data-path and control-path bugs.  
\Omit{
\verifox also outperforms \klee for most configurations in
Scenario~A and Scenario~C.  In Scenario~B, \klee marginally wins over
\verifox by a fraction of a second.  \cbmcv performs poorly for most
scenarios, which is mostly attributed to higher loop bound ($>500$) in the
SW, thereby causing \cbmcv to get stuck in unwinding the program netlist
model.  

Note that both \klee and \verifox were run with the same configurations --
depth-first exploration strategy, eager infeasibility check, and incremental
solving.  \klee uses the STP theorem prover, whereas \verifox uses
MiniSAT~2.2.0 in the backend.
}
\Omit
{Besides SAT backend, \verifox also have a SMT backend. We ran our 
experiments with Z3 and observe higher verification times for Z3 compared 
to MiniSAT. So, we do not report the verification with SMT solvers.   }

\begin{table}
\begin{center}
{
\begin{scriptsize}
\begin{tabular}{|l|l|l|l|l|l|l|}
\hline
  & & \multicolumn{1}{c|}{Monolithic} & \multicolumn{3}{c|}{Path-based} &
  Verification \\ 
\cline{3-6}
  Scenario & Bound & \multicolumn{1}{c|}{HW-CBMC} & \multicolumn{3}{c|}{\verifox} &
  Results \\ 
\cline{3-7}
      &       &  Bit-level & Total/Feasible & \%-age & $\mathcal{M}_{seq}$ &
      Safe/Unsafe \\
  \cline{3-3}\cline{6-6}
      &       &   Time     & Paths & Pruning & Time &  \\
\cline{1-7}      
\multicolumn{7}{|c|}{non-deterministic data but deterministic control (Scenario A)} \\ \hline
  transmit (32) & 250 & 15.02 & 247104/224 & 99.90 & \textbf{1}.\textbf{13} &
  Safe $(\psi_{c})$ \\ 
transmit (64) & 500 & 23.87 & 247104/324 & 99.86 &
  \textbf{1}.\textbf{61}  & Safe $(\psi_{t})$ \\ \hline 
\multicolumn{7}{|c|}{non-deterministic data and non-deterministic control
(Scenario B)} \\ \hline
  trans\_intr (32) & 250 & 14.86 & 247104/295 & 99.88 & \textbf{1}.\textbf{49} &
  Safe $(\psi_{t})$ \\
  trans\_intr (64) & 500 & 24.14 & 247104/362 & 99.85 & \textbf{1}.\textbf{81}
  & Safe $(\psi_{t})$ \\ \hline
\multicolumn{7}{|c|}{non-deterministic data and non-deterministic control
(scenario C)} \\ \hline
  loopback (8)  & 230 & 52.06 & 247104/354 & 99.85 & \textbf{3}.\textbf{95} &
  Safe $(\psi_{t})$ \\ 
  loopback (16) & 500 & 122.12 & 247104/690 & 99.72 & \textbf{12}.\textbf{89} &
  Safe $(\psi_{c})$ \\ 
  loopback (32) & 650 & 170.62 & 247104/1282 & 99.48 & \textbf{21}.\textbf{85} &
  Safe $(\psi_{c})$ \\ 
  loopback (64) & 1300 & 409.71 & 247104/2566 & 98.96 & \textbf{62}.\textbf{31}
  & Safe $(\psi_{t})$ \\ \hline
\multicolumn{7}{|c|}{detecting data-path bugs in transmission mode w/o interrupt} \\ \hline
  transmit (64) & 520 & 28.43 & 247104/324 & 99.86 &
  \textbf{1}.\textbf{12} & Unsafe $(\psi_{t})$ \\ \hline
\multicolumn{7}{|c|}{detecting control bugs with interrupt enabled} \\ \hline
transmit (64) & 520 & 31.35 & 247104/362 & 99.85 &
  \textbf{1}.\textbf{05} & Unsafe $(\psi_{c})$ \\ \hline
\multicolumn{7}{|c|}{detecting control bugs in loopback mode} \\ \hline
  loopback (64) & 1300 & 443.15 & 247104/2566 & 98.96 & \textbf{62}.\textbf{34}
  & Unsafe $(\psi_{t})$ \\ \hline
\end{tabular}
\end{scriptsize}
}
\end{center}
  \caption{Verification of UART (All time in seconds)
\label{table:safe}}
\end{table}

\subsection{HW/SW Co-verification of System-on-Chip}

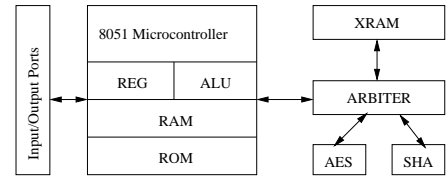
\begin{figure}[t]
\scalebox{.50}{\import{figures/}{soc.pspdftex}}
  \caption{SoC Design obtained from~\cite{DBLP:conf/fmcad/SubramanyanVRM15}
\label{fig:soc}}
\end{figure}

\textbf{\emph{About the SoC:}}
We obtained an open source System-on-Chip design
from~\cite{DBLP:conf/fmcad/SubramanyanVRM15}.  It consists of an 8051
micro-controller, a memory arbiter, an external memory (XRAM) and
cryptographic accelerators, as shown in Figure~\ref{fig:soc}.  
The design statistics for SoC is given in
Table~\ref{table:stats}.  The accelerator implements encryption/decryption 
using the Advanced Encryption Standard (AES). A separate module that interfaces 
the AES to the 8051 micro-controller using a memory-mapped I/O interface.  
The micro-controller communicates 
with the accelerators and the XRAM by reading or writing to 
XRAM addresses.  The arbitration of these module is done by 
the memory arbiter module.
The FW initiates the operation in the SoC by first writing to an
initial memory-mapped register.  The FW implements Linux-style $inb()$
and $outb()$ functions calls, which are used to communicate with the HW
ports.  The FW writes a sequence of non-deterministic data to the XRAM
port and then reads the data from the same port.  The cryptographic
accelerators use direct memory access to fetch the data from the external
memory.  The completion of the operation is determined by polling the
appropriate memory-mapped registers in the FW. \\
\begin{table}
\begin{center}
{
\begin{scriptsize}
\begin{tabular}{|l|l|l|l|l|}
\hline
  & Bound & \multicolumn{1}{c|}{Monolithic} & \multicolumn{1}{c|}{Path-based} &
  Verification \\ 
\cline{3-4}
  Scenario &  & \multicolumn{1}{c|}{\hwcbmcv} & \verifox & Result \\ 
\cline{3-5}
      &       & Bit-level Netlist & $\mathcal{M}_{seq}$ & Safe/Unsafe \\
\cline{1-5}
      &       &   Time (seconds)       & Time (seconds) &  \\
\cline{1-5}      
\multicolumn{5}{|c|}{non-deterministic data and non-deterministic control} \\ \hline
  data\_transfer & 20 & 86.18 & \textbf{17}.\textbf{42} & Safe $(\psi_{t})$ \\
  \hline
  AES\_feedback & 30 & 102.92 & \textbf{56}.\textbf{29} & Safe $(\psi_{c})$ \\ 
\hline
\multicolumn{5}{|c|}{non-deterministic data and non-deterministic control} \\ \hline
  write\_XRAM & 20 & 92.63 & \textbf{14}.\textbf{78} & Unsafe $(\psi_{c})$ \\ 
\hline
  DMA & 32 & 128.63 & \textbf{68}.\textbf{19} & Safe $(\psi_{t})$ \\ 
\hline
\end{tabular}
\end{scriptsize}
}
\end{center}
  \caption{Verification times for SoC Design (All time in Seconds)
\label{table:SoC}}
\end{table}
\textbf{\emph{Discussion of Result:}} 
Table~\ref{table:SoC} gives the runtimes for the bounded safety proofs of 
the SoC design.  Column~1 gives the name of the scenario, 
column~2 report the maximum loop unroll bound of the FW.  Column~3-5 
present the verification runtimes using \hwcbmcv, \verifox and the 
verification outcome respectively.
The result in Table~\ref{table:SoC} shows that \verifox is approximately
$2\times$ faster than \hwcbmcv for proving safety.  For detecting bugs, 
the speedup is $6\times$ for \verifox over \hwcbmcv.  
In the case of SoC scenario (data\_transfer), the SW exercises only the 
micro-controller and transfer sequence of bytes to the XRAM port bypassing 
peripherals connected to other ports such as hardware accelerator.  
This scenario allows path-based symbolic execution engine in \verifox to prune the 
logic for the accelerator and generate only relevant verification conditions for 
the micro-controller and XRAM.  
It is important to note that forward symbolic execution without these optimizations 
timeout for all the benchmarks.
\Omit {The reason for better scalability of \verifox compared to \hwcbmcv is attributed 
to the path-based exploration of the sequential co-verification model using eager 
path pruning, property-guided slicing and incremental SAT solving.  These optimizations 
allow \verifox to automatically mine only relevant scenario/transaction pairs, which 
leads to simple and concise SAT formulas that can be easily verified using state-of-the-art 
SAT solvers.} \\
We verifies a total of~19 properties for the SoC co-design.  Due to space
limitations, we report~4 representative properties in Table~\ref{table:SoC}. 
We check whether the acknowledgement for data transmission and data
reception arrives from the micro-controller in the correct cycle.  We verify
whether the non-deterministic data transmitted through $outb()$ is the same
as the data received through $inb()$.  We also verify that reading/writing
to the appropriate memory-mapped registers produce the correct result during
the data transmission phase.  We found one critical \emph{control bug} in
the SoC design.  The bug is manifested when memory arbiter hardware wrongly
arbitrates the port selection thereby forcing the write strobe for the
external RAM to be LOW.  This violates the data transfer protocol in the SoC
design.
%
%
\section{Limitations of Proposed Approach}
%
The primary motivation for constructing a sequential single threaded unified 
co-verification model is to avoid enumerating exponential number of interleavings 
between the HW and SW threads.  We have shown that this is extremely beneficial for 
co-designs that exhibits producer-consumer relationship.  Such interaction pattern 
is prevalent in many practical co-designs~\cite{polig2014micro,polig2014fpl,giefers2015accelerating,hvc}. 
\Omit{
In particular, our path-based algorithm in \verifox is able to suitably exploit 
this interaction behavior in a sequential co-verification model through symbolic 
execution by generating relevant scenario/transaction pairs.
} 
However, the proposed verification approach is not applicable for co-designs that 
exhibits true concurrency~\cite{codesign}, that is, when a SW and its interacting 
HW threads do not exhibit producer-consumer relationship.  In this case, it is imperative 
to consider all possible interleavings between participating threads in an efficient manner.

%% file: figures/soc.pspdftex
\begin{picture}(0,0)%
\includegraphics{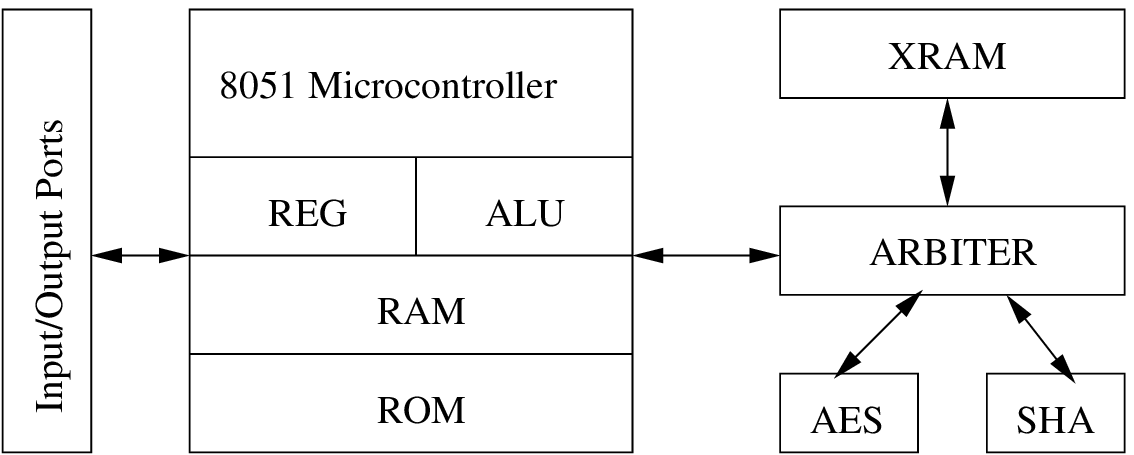}%
\end{picture}%
\setlength{\unitlength}{4144sp}%
\begingroup\makeatletter\ifx\SetFigFont\undefined%
\gdef\SetFigFont#1#2#3#4#5{%
  \reset@font\fontsize{#1}{#2pt}%
  \fontfamily{#3}\fontseries{#4}\fontshape{#5}%
  \selectfont}%
\fi\endgroup%
\begin{picture}(5154,2049)(709,-4573)
\end{picture}%

%% file: background.tex
\section{Related work}

Previous work~\cite{codes14,codes15,fmcad13,memocode06} for 
co-verification have addressed the problem at the pre-RTL phase. 
However, we address the co-veri\-fi\-cation problem at the post-RTL
phase~\cite{fase10,vlsid13} where a key risk is divergence of the 
HW RTL from the behavior expected by the SW. 
Generating a unified co-verification model is a well-known 
technique in HW/SW co-verification.  Notably, Kurshan et al.~modeled HW and 
SW using finite state machines~\cite{fmsd02}, Monniaux in~\cite{emsoft07} modeled
HW and SW as C programs which are formally pushdown systems (PDS), 
Li et al.~in~\cite{fase10} used Buchi Automata to abstractly
model a hardware and PDS to abstractly model a software to generate a unified SW-HW model, 
called Buchi Pushdown System ({\em BPDS}).  In this paper, we construct a unified 
sequential co-verification model in C language.

\Omit{
A system is comprised of a set of concurrent FW and HW models, which
interleave asynchronously.  Malik et al.~in~\cite{hvc} showed 
that these patterns are meaningful in further
settings -- a Linux device driver interacting with x86 QEMU emulator code or
a Rockbox firmware interacting with an iAudio X5 device.  Practical industrial 
co-designs, such as IBM Text Accelerator co-design~\cite{polig2014micro,polig2014fpl} 
and IBM coherent FFT coprocessor~\cite{giefers2015accelerating} also exhibits
producer-consumer interaction pattern. 
}
Common practice in industry for system-level co-verification is to 
either use emulators/accelerators or Instruction Set Simulators 
(ISS)~\cite{coverif-book}.  However, no rigorous formal verification 
effort is performed at the post-RTL phase to ensure the validity of the
SW-HW interactions.  

\Omit{
The concept of symbolic execution~\cite{DBLP:journals/tse/Clarke76,
DBLP:conf/pldi/GodefroidKS05, DBLP:conf/osdi/CadarDE08} is prevalent in the
software domain for automated test generation as well as bug finding.  This
technique is different from the symbolic simulation techniques that are used
in the hardware domain.  In this paper, we apply path-wise symbolic execution
to mine relevant scenario/transaction pairs for effective co-verification of 
a unified co-verification model.  
}

%% file: conclusion.tex
\section{Concluding Remarks}
%
In this paper, we presented a formal HW/SW co-verification tool called \verifox.   
In a typical HW/SW co-design, the software only exercises a fragment
of the HW state-space.  This renders many interactions between HW and
SW modules infeasible.  Our general observation is that the bounded 
model checking technique in \hwcbmcv cannot prune irrelevant logic, 
and hence generates formulas that are extremely difficult to solve with a SAT/SMT solver.
In contrast, the path-based exploration strategy in \verifox is
able to automatically prune design logic with respect to a given 
configuration (scenario), owing to domain-specific optimizations such as eager 
path pruning combined with incremental SAT solving and property-guided slicing.
Our experiments show that \verifox is on average $5\times$ faster than \hwcbmcv 
for proving safety as well as for finding critical bugs.  
In the future, we plan to extend \verifox to support HW/SW co-designs 
that exhibit further interaction patterns as well as implement efficient 
domain-specific path-merging techniques.